\documentclass[aps,rmp,preprint,groupedaddress]{revtex4}

\usepackage{graphicx}

\begin{document}


\title{Discovering the Value of Multidisciplinary Approaches to Research: Insights from a Sabbatical}


\author{Frank W. Bentrem}
\email[]{sciguy137@yahoo.com}
\thanks{Present address: Physics Department, Northwestern College, Orange City, Iowa 51041 USA}
\affiliation{Marine Geosciences Division, Naval Research Laboratory, Stennis Space Center, Mississippi 39525 USA}


\date{December 2009}

\begin{abstract}
In this informal report, I outline my research efforts, collaborations, and other experiences while participating in the Naval Research Laboratory (NRL)'s Advanced Graduate Research Program (AGRP), aka Sabbatical, from October 2008 through September 2009. This report is in no way intended to present the technical details of the various research projects, but rather a broad overview of the small ways my efforts may have contributed to ongoing research. I wish to convey to the reader the value of multidisciplinary approaches to scientific research and how the AGRP facilitates these opportunities.

Disclaimer: The views expressed in this article are those of the author and do not represent opinion or policy of the US Navy or Department of Defense. 
\end{abstract}

\pacs{}

\maketitle
\tableofcontents
\section{Why Pursue the AGRP?}
What is the purpose of the Advanced Graduate Research Program (AGRP) and why would a researcher at the Naval Research Laboratory (NRL) consider applying for the program? The answer to the first question can be found in NRL's Human Resources instruction document \cite{nrlhroinst}:

\begin{quote}
It is NRL policy to maintain a highly competent corps of professional personnel by providing opportunities for employees to keep abreast of advances in
their fields. The purpose of the Advanced Graduate Research Program is to permit selected employees to pursue collaborative research in their own or related fields on a full-time basis.
\end{quote}

\noindent 
To answer the latter question, I will provide my personal motivation for embarking on this year-long program.

\subsection{Polymer Physics Background}
\begin{figure}
\includegraphics*[width=3.375in]{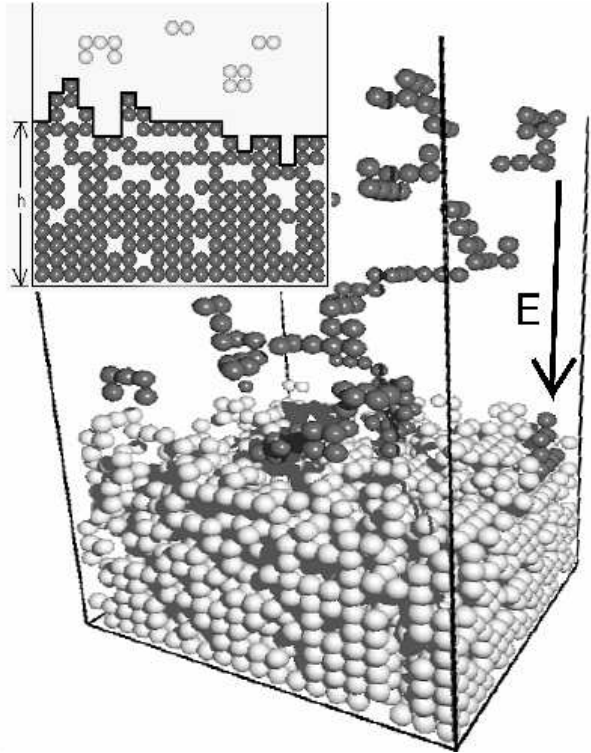}
\caption{\label{fig:1} Simulation snapshot \cite{bentrem00} for polymer electrocoating with uniform electric field $E$. Inset shows surface roughness from a cross section with thickness $h$.}
\end{figure}

My career with NRL began in May 2000 while a part-time graduate student in computational polymer physics. Although my efforts with NRL were focused on acoustic imagery \cite{bentrem06,bentrem09} and sediment classification \cite{brown01,bentrem02,bentrem06,harris08}, I completed my doctoral research in simulations for polymer electrocoatings \cite{bentrem00,bentrem02a,bentrem02b,bentrem05}. (See Fig.~\ref{fig:1}). In recent years, I have searched for ways to use my background and expertise in molecular simulations to best serve the missions of both the Marine Geosciences Division and the laboratory as a whole.

\subsection{Role of Polymers in Marine Sediments}
Much progress is being made towards understanding the physical properties of geologic sediments in terms of the chemical constituents. However, much remains to be understood regarding the influence of organic matter on mechanical and electrical properties of marine sediments. In particular, how do the polymer components (polysaccharides, biomolecules, etc.) of organic matter affect the flocculation, aggregation, and shear strength of muddy sediment on the seafloor? The Marine Geosciences division is increasing efforts to understand these important issues, which impact Navy interests, such as trafficability, mine burial, and beach morphology, to name a few.
\subsection{Opportunities in Polymer Science}

\subsubsection{Tulane's PolyRMC}
In August 2007, the physics department at Tulane University announced the opening of the Center for Polymer Reaction Monitoring and Characterization (PolyRMC) with the mission ``To be the world's premier center for R\&D in polymerization reaction monitoring.'' Since Tulane is a mere 45-minute drive from Stennis Space Center, I soon began interacting with the Center's Director and Assistant Director, Professors Wayne Reed and Alina Alb, to discover potential topics for collaboration. 

\subsubsection{NRL-DC's CBMSE}
About the same time (summer 2007) I became aware of NRL's Center for Bio/Molecular Science and Engineering (CBMSE) and their interests in the theory and simulations for polymer systems. In discussions with some CBMSE researchers, we identified areas where my expertise could complement the work they were doing. I was named as an investigator on a proposal to the NRL Nanoscience Institute. The proposal for an innovative type of body armor was one of fifteen selected for presentation before the nanoscience committee, however, it was not selected to go before the Research Advisory Council.

\subsection{Broad Support}
Although armed with the background and opportunities for polymer research, I realized that for substantial impact to NRL, I would need to arrange for intensive study and research in the most up-to-date applications and techniques both in polymer physics and materials science in general. The AGRP seemed to fit the bill precisely, and I began to discuss this possibility with my supervisors at NRL. Only the Assistant Director of Research (ADOR) need write a letter of recommendation for the AGRP candidate, yet I received tremendous support within my division with first-level supervisor, branch head, and division superintendent all kindly recommending my participation in the program in addition to that of the ADOR. Also, I received letters of invitation from Tulane University and NRL's CBMSE. In spring 2008, to my great delight, I was notified that I had been selected for the AGRP.

\section{Polymer Physics with the Tulane Green Wave}

\subsection{Hello, Mate!}

On the morning of October 1, 2008, I arrived at the PolyRMC and received a staff ID card, a shared office, and a brand-new computer. My office mate, Tomasz Kreft, was in the last year of his doctoral work, and made me feel most welcome. His newest experimental results in polymerization kinetics led me to alter my original computational research plans. In April 2009, after his dissertation defense, Tomasz accepted my invitation to present his research as a seminar at NRL-Stennis.

\subsection{Delivering Drugs (Legally)}
One particular Ph.D. student, Colin McFaul, had been researching a thermo-responsive polymer, that is, one that changes its properties with a change in temperature. Colin demonstrated a sharp change in the polymer size for a Poly(N-isopropylacrylamide), or simply PNIPAm, solution when temperature crossed $33^\circ$C. Since this is very close to body temperature ($37^\circ$C), there is much interest in applications for drug delivery. At temperatures lower than $33^\circ$C, the swollen PNIPAm polymer chains can exist as a gel which encapsulates a given drug. After ingesting the PNIPAm capsule, the warming polymer collapses and the gel dissolves releasing its contents. The mechanism for (and hence the ability to control) this collapse is still under investigation. Much of my research at Tulane focused on developing a computer simulation technique for understanding this temperature dependent transition. 

\subsection{Building a Better Mousetrap--Polymer Simulations}
\label{sec:mousetrap}
\begin{figure}
\includegraphics*[width=5in]{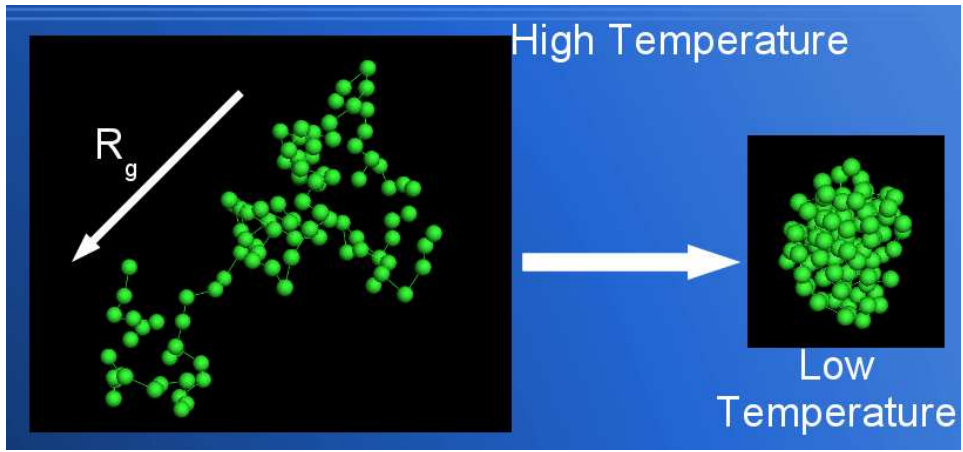}
\caption{\label{fig:2} Collapse of a self-attracting polymer chain. Chain at high temperature (left) has a large radius of gyration, but at low temperature (right) the same chain has collapsed with a small radius of gyration.}
\end{figure}

The bond-fluctuation model (BFM) is a computationally efficient simulation method for researching polymer systems at time and length scales not accessible to other methods. After appropriately assigning the interaction potentials, the BFM has proved successful in reproducing much of the measured polymer structures and dynamics. However, accuracy from the BFM suffers in the highly constrained geometries of dense polymer melts and tightly collapsed chains (as with collapsed PNIPAm globules). I developed an enhancement to the BFM which greatly increased the flexibility in the polymer chains, yet retained the computational efficiency. The resulting polymer coil-globule transition demonstrates the compact collapse (shown in Fig.~\ref{fig:2}) expected for PNIPAm. This enhanced BFM provides broader capability for the simulation of polymer systems. Colin was taking a computational physics course and asked me for a project idea. He helped me modify the enhanced BFM to simulate a dense polymer melt and analyze the dynamics.

\subsection{A Distinguished Gentleman}
On one occasion, while briefing Wayne and Alina (the PolyRMC directors) on my recent progress at the hallway blackboard, a senior gentleman approached and entered the discussion. His insights into the polymer coil-globule transition research were most welcome. He later visited my office for further discussion on my current simulation technique, and mentioned that years earlier he had worked on polymer simulations with Walter Stockmayer--a well known pioneer in polymer simulations whose research was a forerunner to my current simulation method. A friendly and engaging man, he had taken a keen interest in my work. Later, I searched the internet for his name, Dr. Hyuk Yu. It turns out that he is the Walter H. Stockmayer Professor of Chemistry at the University of Wisconsin-Madison, an American Physical Society (APS) fellow, and had received the APS High Polymer Physics Prize! We continue to correspond and plan to meet at the 2010 APS March meeting.  

\section{Researching Soft Matter at NRL-DC}
The second phase of my sabbatical was spent at the Center for Bio/Molecular Science and Engineering in Washington, DC (NRL). I drove to DC with my wife, Amelia, and 3-year-old daughter, Abby, in mid-May 2009 to settle into a modest furnished apartment for a 3-month stay. 

\subsection{Micro- and Nano-scale Plastic Spaghetti}
\begin{figure}
\includegraphics*[width=3.375in]{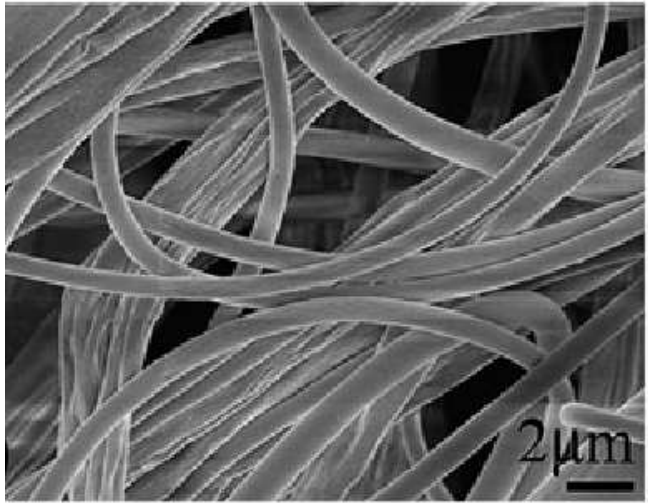}
\caption{\label{fig:3} Polymerized microfibers under magnification \cite{thangawng09}.}
\end{figure}

I continued my work with the center's senior scientist, and the ``nano-Play-Doh group'' on the polymerization kinetics in microfluid channels. NRL has developed a microfluidic system with remarkable control of the shape of the microfluid channels. Polymerizing the material in the microfluid channel produces polymer microfibers (Fig.~\ref{fig:3}) of predetermined shapes. A potential application for the microfibers is in producing high-strength materials such as body armor. My results for the kinetics of the photopolymerization in the microfluid device were intended to be used as a guide for the design of the apparatus in producing micro- and nano-scale fibers. 

\subsection{Artificial Muscles}
Another group I worked with at CBMSE was the liquid-crystal (LC) group. I became interested in an electroclinic LC experiment that they had recently performed. An electroclinic LC is one for which the aligned LC components tilt in either direction under the influence of a switching electric field. The tilting causes the electroclinic LC to contract so that it behaves as an artificial muscle controlled by the applied electric field. Experiments had been performed both with the ordinary electroclinic LC and an electroclinic LC elastomer, where polymer backbones attached to the aligned LC components. The data showed a puzzling difference between the two experiments presumed to be caused by the polymer dynamics. Since I had experience in this area, I performed an analysis on the tilt dynamics and related this to the theory of dynamics for entangled polymer. Interaction among the LC group members is quite active and I had regular discussions and input from various group members. Key parts of this work were included in a recent paper \cite{spillmann10}.  

\begin{figure}
\includegraphics*[width=3.375in]{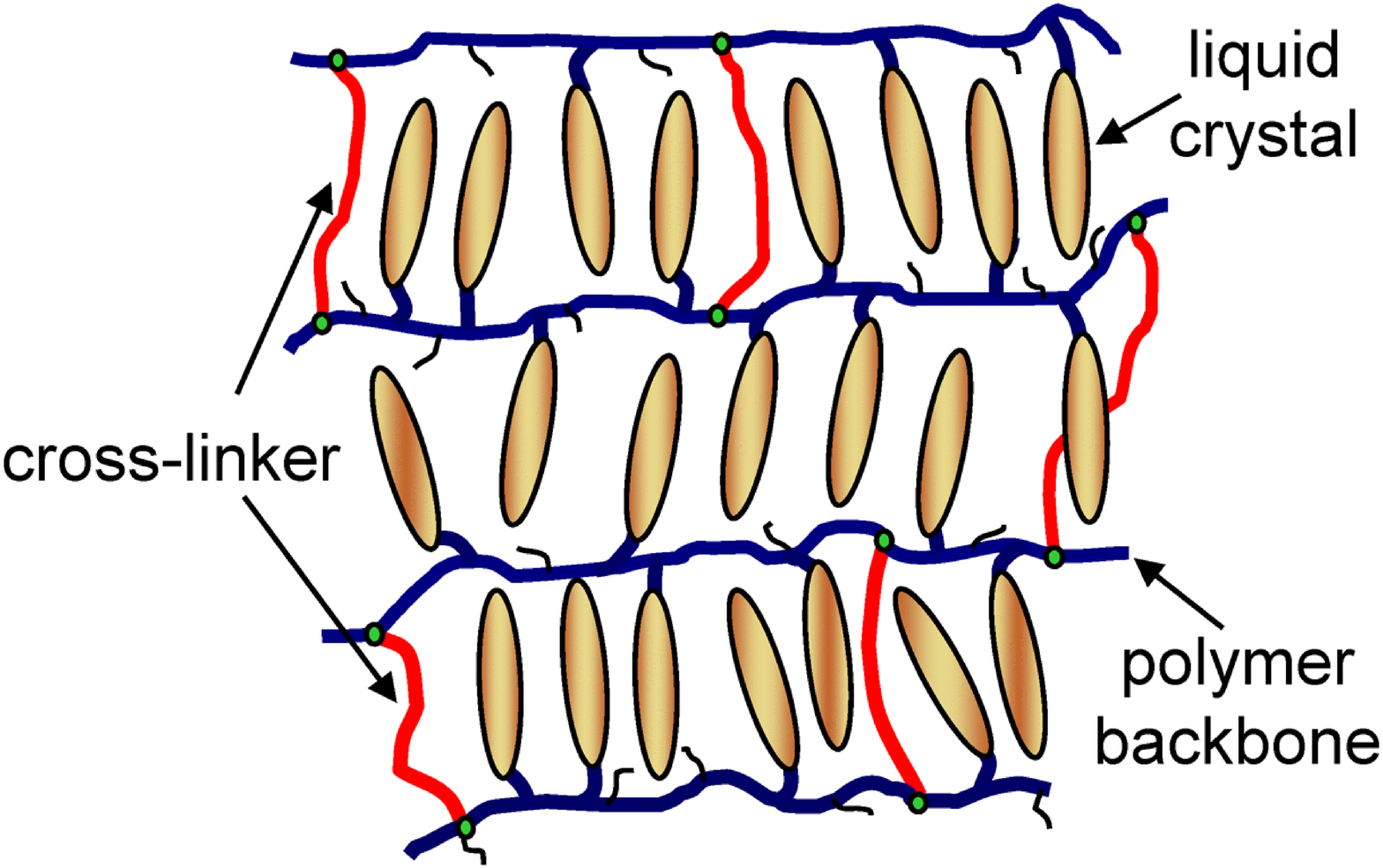}
\caption{\label{fig:4} An electroclinic LC elastomer with cross-linking \cite{spillmann10}.}
\end{figure}

\section{New Directions with Nano-material}

\subsection{It's a Bird, It's a Plane, It's Super-Carbon!}

\begin{figure}
\includegraphics*[width=3.375in]{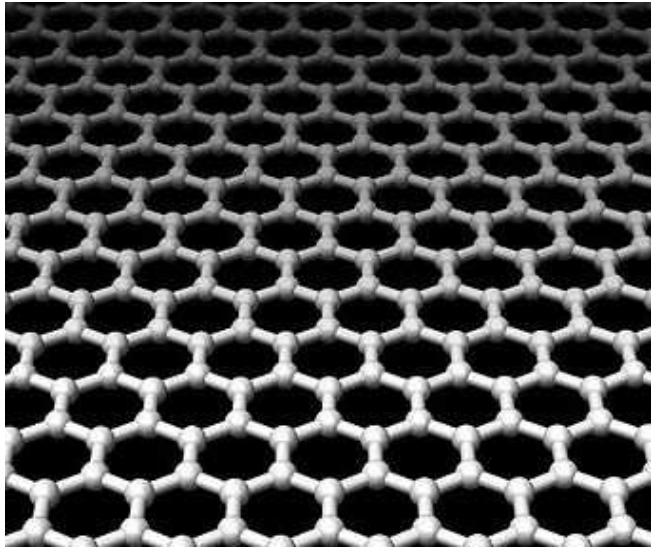}
\caption{\label{fig:5} A flat graphene sheet \cite{wikipedia} highlighting the characteristic honeycomb pattern.}
\end{figure}

While listening to some short proposal presentations at CBMSE, my ears perked up at a proposal for experimental research on graphene. Graphene is a 1-atom thick, pure-carbon sheet (See Fig.\ref{fig:5}) with extraordinary electronic, mechanical, and optical properties. It is the strongest material ever tested (200 times stronger than steel), the most sensitive chemical sensor possible (shown to detect single molecules), and the thinnest possible membrane (1-atom thick sheet). Assumed not to exist (based on prior theory), researchers were surprised to discover graphene sheets \cite{novoselov04}, and, due to its unique properties and vast assortment of potential applications (particularly in carbon-based nano-electronics and bio/chemical sensing), it has become one of the ``hottest'' topics in physics. One of the great challenges in using graphene as a semiconductor in nano-electronics and bio/chemical sensing is understanding and controlling the ripples that spontaneously form throughout the sheets, which can be seen with electron microscopy.   

\subsection{Hey, We Can Do That}

I imagined extending the (1-dimensional) polymer model from Sec.~\ref{sec:mousetrap} to a 2-dimensional graphene model. Indeed, my doctoral research adviser has already researched clay platelets using a similar technique. The key to modeling graphene would be in using a lattice-in-a-lattice approach to access the fine (nano-)scales appropriate to the carbon bonds. This ultra-efficient technique should allow graphene-ripple simulations at unprecedented time and length scales. The CBMSE researcher (whose proposal sparked my interest in graphene simulations) and I began discussing the possibility for me to lead a 6.1 New Start proposal with the goal of understanding graphene ripples and the adsorption of biomolecules onto graphene sheets with the potential application to marine biosensing.

\section{Final Thoughts}

\subsection{Multidisciplinary Magic}

Prior to my sabbatical year, I had already experienced the benefits of multidisciplinary approaches to research by applying my skills in statistical physics to problems in underwater acoustics \cite{bentrem02,bentrem06,bentrem09}. My sabbatical experience, however, amplified my appreciation for multidisciplinary and collaborative science. It was a privilege to work with the many diverse, enthusiastic, and creative researchers at both Tulane and NRL-DC. I agree wholeheartedly with the following statement by our recently retired distinguished NRL colleague.

\begin{quote}
One of the most important considerations concerning the support of research with public funds is the relation of basic research to applied research or technology. Technology flourishes when it is stimulated by the results of basic research. To illustrate this, refer to a statement made during the 1950's when the attack on poliomyelitus was bearing fruit: ``With technology one can build the best iron lung in the world, but from basic research you have a vaccine''. The administration in charge of research at the NRL clearly understands the relationship between basic research and technology.
\end{quote}

\indent Dr. Jerome Karle\\
\indent    Nobel Laureate\\
\indent    Chief Scientist in NRL's Lab for Structure of Matter

\subsection{Bean Counting}

As a direct result of my efforts in the AGRP, I have:

\begin{itemize}
\item{} 1 accepted refereed journal article \cite{bentrem09} (sole author)
\item{} 1 submitted journal article (coauthor)
\item{} 1 manuscript in progress (lead author)
\item{} 1 international conference presentation (coauthor)
\item{} 2 future international conference presentations (March 2010, lead author)
\item{} 1 Provisional Patent Application filed and approved by NRL's Invention Evaluation Board \cite{patent}
\item{} 1 DTRA proposal submitted (not funded)
\item{} 1 planned proposal
\item{} Enhanced polymer simulation model ideally suited for the Marine Geosciences Division's ARI (starting in FY10).
\end{itemize}

My partaking in the AGRP (sabbatical) was time and effort well spent. As the old adage goes, ``A wood chopper never wastes time when he takes the time to sharpen his ax''. By broadening and deepening my skills in science, math, and computation, I believe I am well-positioned to creatively solve the scientific and engineering problems I will encounter at NRL.



\bibliography{sabbatical2}

\end{document}